\documentclass[fleqn,11pt,twoside]{article}
\usepackage{amsthm}
\usepackage{amsmath, amscd}%[fleqn,intlimits]
\usepackage{graphicx}
\usepackage{lscape}
\usepackage{amsmath}%[fleqn,intlimits]
\usepackage{graphicx}
\usepackage{latexsym}
\usepackage[all]{xy}
\makeatletter

\newcommand{\Name}[1]{\begin{flushleft}
                       \LARGE \bf #1
                       \end{flushleft}\vspace{-3mm}}

\newcommand{\Author}[1]{\begin{flushleft}
                       \it #1 \end{flushleft}}

\newcommand{\Address}[1]{\begin{flushleft}
                       \it #1 \end{flushleft}}

\newcommand{\FirstPageHead}[5]{
\begin{flushleft}
\raisebox{8mm}[0pt][0pt]
{\footnotesize \sf
\parbox{150mm}{ \qquad
 #1 #2 #3 
#4\hfill {\sc #5}}}\vspace{-13mm}
\end{flushleft}}

\newcommand{\evenhead}{Author \ name}
\newcommand{\oddhead}{Article \ name}

\renewcommand{\@evenhead}{
\hspace*{-3pt}\raisebox{-15pt}[\headheight][0pt]{\vbox{\hbox to \textwidth
{\thepage \hfil \evenhead}\vskip4pt \hrule}}}
\renewcommand{\@oddhead}{
\hspace*{-3pt}\raisebox{-15pt}[\headheight][0pt]{\vbox{\hbox to \textwidth
{\oddhead \hfil \thepage}\vskip4pt\hrule}}}
\renewcommand{\@evenfoot}{}
\renewcommand{\@oddfoot}{}

\long\def\@makecaption#1#2{%
  \vskip\abovecaptionskip
  \sbox\@tempboxa{\small \textbf{#1.}\ \ #2}%
  \ifdim \wd\@tempboxa >\hsize
    {\small \textbf{#1.}\ \ #2}\par
  \else
    \global \@minipagefalse
    \hb@xt@\hsize{\hfil\box\@tempboxa\hfil}%
  \fi
  \vskip\belowcaptionskip}

\newcommand{\JNMPnumberwithin}[3][\arabic]{%
  \@ifundefined{c@#2}{\@nocounterr{#2}}{%
    \@ifundefined{c@#3}{\@nocnterr{#3}}{%
      \@addtoreset{#2}{#3}%
      \@xp\xdef\csname the#2\endcsname{%
        \@xp\@nx\csname the#3\endcsname .\@nx#1{#2}}}}%
}

\newcommand{\resetfootnoterule} {
  \renewcommand\footnoterule{%
  \kern-3\p@
  \hrule\@width.4\columnwidth
  \kern2.6\p@}
}

\renewcommand{\footnoterule}{}

\newcommand{\be}{\begin{equation}}
\newcommand{\ee}{\end{equation}}
\newcommand{\ba}{\hspace*{-5pt}\begin{array}}
\newcommand{\ea}{\end{array}}
\newcommand{\p}{\partial}

\makeatother
\numberwithin{equation}{section}
\theoremstyle{definition}

\renewcommand{\ba}{\begin{array}}
\renewcommand{\ea}{\end{array}}
\newcommand{\beg}{\begin{eqnarray}}
\newcommand{\eeq}{\end{eqnarray}}
\newcommand{\bg}{\begin{eqnarray*}}

\newcommand{\ed}{\end{eqnarray*}}

\renewcommand{\p}{\partial} 
 
\newcommand{\notlhd}{\lhd\kern-.8em{/}\ } 
\newcommand{\notexist}{\ \exists\kern-.5em{\raise.1em\hbox{/}}\ }

\newcommand{\pde}[2]{\frac{\p #1}{\p #2}}

\newcommand{\inp}{{\mbox{\vbox{\hrule width0ex\hbox{\vrule
 height0ex\kern3.8pt
\vbox{\kern2.5pt}\kern3.8pt \vrule height1.6ex}
\hrule width1.6ex}}}}

\begin{document}

%\Large

\renewcommand{\evenhead}{}
\renewcommand{\oddhead}{}

% Title

\thispagestyle{empty}

\begin{flushleft}
\footnotesize \sf
\end{flushleft}

\FirstPageHead{\ }{\ }{\ }
{ }{{
{ }}}
%\copyrightnote{2012}{M Euler and N Euler}

\Name{Invariance of the Kaup-Kupershmidt equation and
  triangular auto-B\"acklund transformations\footnote{Communicated by
  F. Calogero}}

\label{firstpage}

%\strut\hfill

%\strut\hfill

%\strut\hfill

\Author{Marianna Euler and Norbert Euler}

%\strut\hfill

%\strut\hfill

\Address{
%\noindent
Department of Engineering Sciences and Mathematics\\ 
Lule\aa\ University of Technology\\
SE-971 87 Lule\aa, Sweden\\
Emails: marianna@ltu.se;  norbert@ltu.se}

%\vspace{1cm}

%\vspace{1cm}

%\strut\vfill

%\pagebreak

\noindent
{\bf Abstract}: 
We report triangular auto-B\"acklund transformations for the solutions
of a fifth-order evolution equation, which is a constraint for 
an invariance condition of the Kaup-Kupershmidt equation 
derived by E. G. Reyes in
his paper titled ``Nonlocal symmetries and the Kaup-Kupershmidt 
equation'' [{\it J. Math. Phys.}  {\bf 46},
073507, 19 pp., 2005]. These auto-B\"acklund transformations can then
be applied to generate solutions of the Kaup-Kupershmidt equation.
We show that triangular auto-B\"acklund transformations result
from a systematic multipotentialisation of the Kupershmidt equation.

%%%%%%%%%%%%%%%%%%%%%%%%

\renewcommand{\theequation}{\arabic{section}.\arabic{equation}}

\section{Introduction}

\noindent
In his paper \cite{Reyes}, Reyes 
reports an invariance of the Kaup-Kupershmidt equation
\begin{gather}
\label{KK-q}
q_t=q_{xxxxx}+5qq_{xxx}+\frac{25}{2}q_xq_{xx}+5q^2q_x,
\end{gather}
by the following

\strut\hfill

\noindent
{\bf Proposition 1:} \cite{Reyes} {\it
The Kaup-Kupershmidt equation, (\ref{KK-q}), is invariant under the
transformation $q\mapsto \bar q$, in which
\begin{gather}
\label{q-bar-eq}
\bar q=q+3\left(\ln B\right)_{xx},
\end{gather}
where the variables $q$ and $B$ are related by
\begin{gather}
\label{q-invariance}
q(x,t)=-\frac{B_{xxx}}{B_x}+\frac{3}{4}\left(\frac{B_{xx}}{B_x}\right)^2
\end{gather}
and $B(x,t)$ is a solution to
\begin{gather}
\label{B-eq}
B_t=B_{xxxxx}-5\frac{B_{xx}B_{xxxx}}{B_x}
-\frac{15}{4}\frac{B_{xxx}^2}{B_x}
+\frac{65}{4}\frac{B_{xxx}B_{xx}^2}{B_x^{2}}
-\frac{135}{16}\frac{B_{xx}^4}{B_x^3}.
\end{gather}
}

\strut\hfill

In the current letter we report a class of 
auto-B\"acklund transformations, known as the
triangular auto-B\"acklund transformations (or
$\bigtriangleup$-auto-B\"acklund transformations \cite{EE-converse}), 
by a systematic multipotentialisation of the equation (\ref{B-eq}).
Triangular auto-B\"acklund transformations have been defined and
demonstrated in our recent paper \cite{EE-converse}.
For details on the multipotentialisation procedure we refer the reader
to our papers \cite{EE-nonlocal} and
\cite{EE-KN}. 
We show that the equation,
\begin{gather}
\label{vv-eq}
v_t=v_{xxxxx}-5\frac{v_{xx}v_{xxxx}}{v_x}+5\frac{v_{xx}^2v_{xxx}}{v_x^2}
\end{gather}
plays a central role in the construction of solutions of the
Kaup-Kupershmidt equation and consequently we find the
general stationary solution of (\ref{vv-eq}) by calculating
its first integrals.

We remark that both equations (\ref{B-eq}) and (\ref{vv-eq}) 
first appeared in a paper by J. Weiss \cite{Weiss}
on the Painlev\'e analysis in the form of singularity manifold
constraints 
for the Kupershmidt equation
(see eq. (3.32) in \cite{Weiss}) and
the Caudrey-Dodd-Gibbon equation (see eq. (3.21) in \cite{Weiss}), 
respectively. For more details on
the
Painlev\'e analysis and singularity manifolds we refer the reader 
to \cite{Steeb-Euler}.
Furthermore, both (\ref{B-eq}) and (\ref{vv-eq}) 
are
known symmetry-integrable equations 
and appear in \cite{M} as part of the list of fifth-order 
semilinear integrable evolution equations (see equations (4.2.12) and (4.2.11)
in \cite{M}). The recursion operators for both equations (\ref{B-eq}) 
and (\ref{vv-eq}) are given in \cite{EE-5th}.

\section{Multipotentialisations and $\bigtriangleup$-auto-B\"acklund
  transformations}
Let 
\begin{gather}
\label{F-u}
u_t=F[u]
\end{gather}
denote an evolution equation in $u$, where $F[u]$ denotes a given
function that depends in 
general on $x$, $t$, $u$ and $x$-derivatives of $u$. 
The procedure to potentialise (\ref{F-u}) in the equation
\begin{gather}
\label{G-v}
v_t=G[v_x]
\end{gather}
is well known (see e.g. \cite{Anco-Bluman-1} and \cite{EE-KN}). 
Equation (\ref{G-v}) is 
  known as the potential equation of (\ref{F-u}).
For the benefit of the reader
and to establish the notation we describe this procedure briefly: The
potentialisation of (\ref{F-u}) in (\ref{G-v}), if it exists, 
is established by a conserved current $\Phi^t[u]$ of (\ref{F-u})
and the relation to the potential variable $v$ is then
\begin{gather}
v_x=\Phi^t[u],
\end{gather}
where 
\begin{gather}
\left.
\vphantom{\frac{DA}{DB}}
D_t\Phi^t[u]+D_x\Phi^x[u]\right|_{u_t=F[u]}=0
\end{gather}
and $\Phi^x$ is the conserved flux of (\ref{F-u}).
Here $D_t$ and $D_x$ are the total $t$- and $x$-derivatives, repectively.
Conserved currents, $\Phi^t[u]$, for (\ref{F-u}) can be obtained
by the relation
\begin{gather}
\Lambda[u]=\hat E[u]\Phi^t[u],
\end{gather}
where $\Lambda$ is an integrating factor (also called multiplier) of
(\ref{F-u})
that can be calculated by the relation
\begin{gather}
\hat E[u]\left( \Lambda[u] u_t-\Lambda[u]F[u]\right)=0.
\end{gather}
Here $\hat E[u]$ is the Euler operator,
\begin{gather*}
\hat E[u]=\pde{\ }{u}-D_x\circ \pde{\ }{u_x}-D_t\circ\pde{\ }{u_t}+D^2_x\circ 
\pde{\ }{u_{2x}}-D_x^3\circ \pde{\ }{u_{3x}}+\cdots.
\end{gather*}
A multipotentialisation of (\ref{F-u}) exists if (\ref{G-v}) can
also be potentialised. A $\bigtriangleup$-auto-B\"acklund
transformation for (\ref{F-u}) exists if (\ref{G-v}) potentialises
back into
the original equation (\ref{F-u}). In fact we have define three
different types of $\bigtriangleup$-auto-B\"acklund
transformations. Details are in \cite{EE-converse}, where several
examples of $\bigtriangleup$-auto-B\"acklund
transformations are given for some third-order and fifth-order
evolution equations, as well as for systems in (1+1) dimensions and
higher dimensional evolution equations.

\strut\hfill

Our starting point is the Kupershmidt equation
\begin{gather}
\label{Kuper}
K_t=K_{xxxxx}+\lambda\left(K_xK_{xxx}+K^2_{xx}\right)
-\frac{\lambda^2}{5}\left(K^2K_{xxx}
+5K_xK_{xx}+K_x^3\right)
+\frac{\lambda^4}{125}K^4K_x
\end{gather}
which potentialises under 
\begin{gather}
U_x=K 
\end{gather}
in the first potential Kupershmidt equation,
\begin{gather}
\label{PK-U}
U_t=U_{xxxxx}+\lambda U_{xx}U_{xxx}-\frac{\lambda^2}{5}\left(
U_x^2U_{xxx}+U_xU_{xx}^2\right)+\left(\frac{\lambda}{5}\right)^4U_x^5,
\end{gather}
and, in turn again potentialises under 
\begin{gather}
\label{Reyes-transf}
u_x=-\frac{5}{2\lambda}\exp\left(-\frac{2\lambda}{5}U\right)
\end{gather}
in a second-potential Kupershmidt equation,
\begin{gather}
\label{Reyes-u}
u_t=u_{xxxxx}-\frac{5u_{xx}u_{xxxx}}{u_x}-\frac{15u_{xxx}^2}{4u_x}
+\frac{65u_{xx}^2u_{xxx}}{u_x^2}-\frac{135u_{xx}^4}{16u_x^3}.
\end{gather}
In addition, the first potential Kupershmidt equation (\ref{PK-U}) also
potentialises in
\begin{gather}
\label{v-eq}
v_t=v_{xxxxx}-5\frac{v_{xx}v_{xxxx}}{v_x}+5\frac{v_{xx}^2v_{xxx}}{v_x^2}
\end{gather}
under 
\begin{gather}
v_x=\frac{5}{\lambda}\exp \left(\frac{\lambda}{5}U\right).
\end{gather}
A connection between the Kupershmidt equation, (\ref{Kuper}),
and the equation (\ref{Reyes-u}) is given by the differential
substitution
\begin{gather}
\label{K-R-connection}
K(x,t)=-\frac{5}{2\lambda}\frac{u_{xx}}{u_x},
\end{gather}
which is obtained by combining $U_x=K$ and
$u_x=-5/(2\lambda)\exp(-2U/(5\lambda)$.

%This multipotentialisation of the Kupershmidt equation can be
%continued beyond (\ref{v-eq}) as shown in Diagram 1 below.

Furthermore, we can connect the first
potential Kupershmidt equation, (\ref{PK-U}), and the 
Kaup-Kupershmidt equation
\begin{gather}
Q_t=Q_{xxxx}+\mu
QQ_{xxx}+\frac{5\mu}{2}Q_xQ_{xx}+\frac{\mu^2}{5}Q^2Q_x
\end{gather}
for arbitrary constant $\mu\neq 0$, by the differential substitution
(given in \cite{M} for the special case, $\mu=5$ and $\lambda=5$)
\begin{gather}
Q=\left(\frac{2\lambda}{\mu}\right)U_{xx}-\left(\frac{\lambda}{5\mu}\right)U_x^2.
\end{gather}

We now focus our attention on equation (\ref{Reyes-u}). For this
equation the most general second-order integrating factors are
\begin{gather}
\Lambda_1[u]=\alpha u_x^{-5/2}u_{xx},\qquad
\Lambda_2[u]=\alpha\left(
uu_x^{-5/2}u_{xx}-2u_x^{-1/2}\right),
\end{gather}
which leads to two potentialisations in (\ref{v-eq})
with $\alpha=-3/4$, namely
\begin{gather}
v_{1,x}=u_x^{-1/2},\qquad
v_{2,x}=uu_x^{-1/2},
\end{gather}
respectively. See the Diagram below.
Furthermore, the most general second-order integrating factors
of equation (\ref{v-eq}) are
\begin{gather}
\Lambda_n[v]=v^n\frac{v_{xx}}{v_x^2}-\frac{n}{2}v^{n-1}\frac{1}{v_x^2},\quad
n=0,1,2,3,4.
\end{gather}
The integrating factor $\Lambda_0$ and $\Lambda_4$ lead to the
potentialisation of equation (\ref{v-eq}) in equation
(\ref{Reyes-u}), namely
\begin{gather}
u_{3,x}=v_x^{-2},\qquad
u_{4,x}=v^4v_x^{-2},
\end{gather}
respectively. See the Diagram below.

\begin{displaymath}
\xymatrix{
&
\mbox{\underline{Diagram:}}\\
\mbox{Eq.(\ref{Reyes-u}) in } u_1
\ar[dr]^{v_x=u_{1,x}^{-1/2}}&\qquad &\mbox{Eq.(\ref{Reyes-u}) in } u_2
\ar[dl]_{v_x=u_2u_{2x}^{-1/2}\ \ }\\
&\framebox{$
v_t=v_{xxxxx}-5v_x^{-1}v_{xx}v_{xxxx}+5v_x^{-2}v_{xx}^2v_{xxx}$}
\ar[dr]_{u_{4,x}=v^4v_x^{-2}\ \ \ \ }\ar[dl]^{u_{3,x}=v_x^{-2}}\\
\mbox{Eq.(\ref{Reyes-u}) in } u_3&\qquad & \mbox{Eq.(\ref{Reyes-u}) in } u_4
}
\end{displaymath}

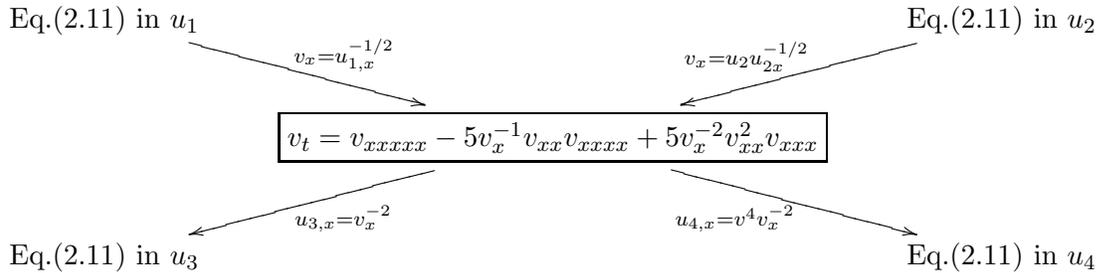
\strut\hfill

\noindent
Combining the above potentialisations for the equation
(\ref{Reyes-u})
and the equation (\ref{v-eq}), leads to the following
$\bigtriangleup$-auto-B\"acklund transformations for these equations:

\strut\hfill

\noindent
{\bf Proposition 2:} {\it
The equation (\ref{Reyes-u}), viz.
\begin{gather*}
%\label{Reyes-eq}
u_t=u_{xxxxx}-5\frac{u_{xx}u_{xxxx}}{u_x}
-\frac{15}{4}\frac{u_{xxx}^2}{u_x}
+\frac{65}{4}\frac{u_{xxx}u_{xx}^2}{u_x^{2}}
-\frac{135}{16}\frac{u_{xx}^4}{u_x^3}
\end{gather*}
admits the following $\bigtriangleup$-auto-B\"acklund transformations:
\begin{subequations}
\begin{gather}
\label{Prop-2-a}
u_{j+1,xx}=u_{j+1,x}
\left[\frac{u_{j,xx}}{u_{j,x}}-2\frac{u_{j,x}}{u_j}\right]
+4\,u_{j+1,x}^{3/4}\left[\frac{u_j^{1/2}}{u_{j,x}^{1/4}}\right],\\[0.3cm]
\label{Prop-2-b}
u_{j+1,xx}=u_{j+1,x}\left[\frac{u_{j,xx}}{u_{j,x}}\right]
+4\, u_{j+1,x}^{3/4}\left[\frac{1}{u_{j,x}^{1/4}}\right],
\end{gather}
\end{subequations}
where
$u_j$ and $u_{j+1}$ satisfy (\ref{Reyes-u}) 
for all natural
numbers $j$.}

\strut\hfill

\noindent
{\bf Proposition 3:} {\it
The equation (\ref{v-eq}), viz.
\begin{gather*}
%\label{v-eq}
v_t=v_{xxxxx}-5\frac{v_{xx}v_{xxxx}}{v_x}+5\frac{v_{xx}^2v_{xxx}}{v_x^2},
\end{gather*}
admits the following $\bigtriangleup$-auto-B\"acklund transformations:
\begin{subequations}
\begin{gather}
%v_{j+1,x}=\frac{v_{j,x}}{v_j^2}\\[0.3cm]
v_{j+1,xx}=v_{j+1,x}\left[\frac{v_{j,xx}}{v_{j,x}}\right]
- \frac{1}{v_{j,x}},\\[0.3cm]
v_{j+1,xx}=v_{j+1,x}\left[\frac{v_{j,xx}}{v_{j,x}}\right]
+ \frac{1}{v_{j,x}},\\[0.3cm]
%v_{j+1,xx}=\frac{1}{2}\,v_{j+1,x}\left[\frac{v_{j,xx}}{v_{j,x}}\right]
%+ \frac{1}{v_{j,x}^{1/2}}\\[0.3cm]
v_{j+1,xx}=v_{j+1,x}\left[\frac{v_{j,xx}}{v_{j,x}}
-2\, \frac{v_{j,x}}{v_j}\right]
+ \frac{v_j^2}{v_{j,x}},
\end{gather}
\end{subequations}
where
$v_j$ and $v_{j+1}$ satisfy (\ref{v-eq}) 
for all natural numbers $j$.}

\strut\hfill

\noindent
{\bf Remark 1:} We note that in addition to the transformations given by 
Proposition 2 and Proposition 3, both (\ref{Reyes-u}) and 
(\ref{v-eq}) admit the symmetry transformation
\begin{gather}
u_{j+1}=-\frac{1}{u_j}+\mbox{constant},\qquad v_{j+1}=-\frac{1}{v_j}+\mbox{constant}
\end{gather}
which are included in the Diagram: combine
$v_x=u_{1,x}^{-1/2}$ with $v_x=u_2u_{2,x}^{-1/2}$, and
$u_x=v_{1,x}^{-2}$ with $u_{x}=v_2^4v_{2,x}^{-2}$, respectively. Note
that these relations were also identified by Weiss (see 
relation (3.37) in \cite{Weiss}).

\strut\hfill

\noindent
{\bf Remark 2:} The $\bigtriangleup$-auto-B\"acklund transformations
(\ref{Prop-2-a}) and (\ref{Prop-2-b})
given in Proposition 2 can be linearised, as those
are in the form of first-order Bernoulli equations under the
substitution $u_{j+1,x}=w_{j+1}$.

\section{Solutions}
In the above Diagram it is clear that equation
(\ref{v-eq}), viz.
\begin{gather*}
%\label{v-eq}
v_t=v_{xxxxx}
-\frac{5v_{xx}v_{xxxx}}{v_x}
+\frac{5v_{xx}^2v_{xxx}}{v_x^2},
\end{gather*}
plays a central role in the connection between the Kupershmidt-,
the Kaup-Kupershmidt and equation (\ref{Reyes-u}). 

We now find the most general stationary solution
for equation (\ref{v-eq})
by calculating the first integrals and the general solution
of the ordinary differential equation
\begin{gather}
\label{v-ode}
v_{xxxxx}
-\frac{5v_{xx}v_{xxxx}}{v_x}
+\frac{5v_{xx}^2v_{xxx}}{v_x^2}=0.
\end{gather}
We find that equation (\ref{v-ode}) admits
five second-order integrating factors, 
$\{\Lambda_0,\Lambda_1,\Lambda_2,\Lambda_3,\Lambda_4\}$:
\begin{gather}
\Lambda_n(x,v,v_x,v_{xx})=a_n\frac{v_{xx}}{v_x^2}-\frac{a_n'}{2}\frac{1}{v_x^2},
\end{gather}
where
\begin{gather*}
a_n(v)=v^n\qquad \mbox{with}\quad n=0,1,2,3,4.
\end{gather*}
Here the primes denote derivatives with respect to $v$.
The corresponding five first integrals are then
\begin{gather*}
I_n=\left(a_n\frac{v_{xx}}{v_x^4}-\frac{1}{2}\frac{a_n'}{v_x^2}\right)v_{xxxx}
-\frac{1}{2}a_n\frac{v_{xxx}^2}{v_x^4}
%\\[0.3cm]
%\qquad
+\left(
\frac{1}{2}a_n'\frac{v_{xx}}{v_x^3}-a_n\frac{v_{xx}^2}{v_x^5}
+\frac{1}{2}\frac{a_n''}{v_x}\right)v_{xxx}\\[0.3cm]
\qquad
-\frac{1}{2}a_n'''v_{xx}+\frac{1}{4}a_n^{(4)}v_x^2,
\qquad n=0,1,2,3,4.
\end{gather*}
Combining $I_0,\ I_1,\ I_2,\ I_3$ and $I_4$, we obtain
\begin{gather*}
v_x=\left(
I_0v^4-4I_1v^3+6I_2v^2-4I_3v+I_4\right)^{1/2},
\end{gather*}
with the condition
\begin{gather*}
I_0I_4-4I_1I_3+3I_2^2=0.
\end{gather*}
So the general solution of (\ref{v-ode})
is given by the quadrature
\begin{gather}
\label{stat_gen}
\int
\left(
I_0v^4-4I_1v^3+6I_2v^2-4I_3v+I_4\right)^{-1/2}dv=x+C,
\end{gather}
where $C$ is a constant of integration.
Since the quadrature (\ref{stat_gen}) contains five independent free constants,
it represents the most general stationary solution of
(\ref{v-eq}). This solution can now be used to
generate nonstationary solutions for (\ref{v-eq}) by the
$\bigtriangleup$-auto-B\"acklund transformations of Proposition 3 and, 
by applying the transformations in the above Diagram, solutions of
(\ref{Reyes-u}) can be constructed.
Then by Proposition 1, the so constructed solutions of (\ref{Reyes-u})
lead to solutions of the Kaup-Kupershmidt equation, as well as solutions
of the Kupershmidt equation by the differential substitution
(\ref{K-R-connection}).

For example, by (\ref{stat_gen}) with $I_0=I_1=I_2=I_4=C=0$ and $I_3=-2$
we obtain the solution
\begin{gather*}
v(x,t)=2x^2
\end{gather*}
for (\ref{v-eq}) and hence the solution
\begin{gather*}
u(x,t)=16x
\end{gather*}
for (\ref{Reyes-u}). Applying now, for example, 
the $\bigtriangleup$-auto-B\"acklund transformation (\ref{Prop-2-a}) leads to the solution
\begin{gather*}
u(x,t)=\frac{1}{7}x^7-\frac{4}{5}c_1x^5+2c_1^2x^3-4c_1^3x-c_1^4x^{-1}-576
c_1t+c_2,
\end{gather*}
where $c_1$ and $c_2$ are arbitrary constants. Appplying again
(\ref{Prop-2-a}), setting $c_1=c_2=0$ for simplicity, we obtain
the following solution for (\ref{Reyes-u}):
\begin{gather*}
u(x,t)=
\frac{1}{5^4\cdot 7\cdot 91}x^{13}
+\frac{36}{5^3\cdot 7}x^8 t
+\frac{2^7\cdot 3^4}{5^2}x^3t^2
-2^{12}\cdot 3^8\cdot 7 x^{-7}t^4
-\frac{2^{10}\cdot 3^6\cdot 7}{5}x^{-2}t^3.
\end{gather*}
Proposition 1 can now be applied with the above solutions of
(\ref{Reyes-u}) (put $u(x,t)\equiv B(x,t)$ in (\ref{q-bar-eq})
and (\ref{q-invariance})) to gain solutions 
for the Kaup-Kupershmidt equation (\ref{KK-q}). Furthermore, solutions of the
Kupershmidt equation,
(\ref{Kuper}), are then given by the differential substitution
(\ref{K-R-connection}).

\section*{Acknowledgements}
We thank the anonymous referees for their constructive
and critical suggestions which led to the improvement of this letter,
and the Wenner-Grenn Foundations for financial support.

\end{document}